\documentclass[10pt]{iopart}

\usepackage[dvips]{graphicx}

\begin{document}

\title{Cosmic ray strangelets}

\author{Jes Madsen}

\address{Department of Physics and Astronomy, University of Aarhus,
DK-8000 {\AA}rhus C, Denmark}

\ead{jesm@phys.au.dk}

\begin{abstract}
Searching for strangelets in cosmic rays may be the best way to test the
possible stability of strange quark matter. I review calculations of the
astrophysical strangelet flux in the GV--TV rigidity range,
which will be investigated
from the Alpha Magnetic Spectrometer (AMS-02) on the International
Space Station, and discuss the merits of strangelets as ultra-high
energy cosmic rays at EeV--ZeV energies,
beyond the Greisen-Zatsepin-Kuzmin cutoff.
I also address some ``counter-arguments'' sometimes raised against 
the possibility of stable strangelets. It will be argued that stability
of strange quark matter remains a viable possibility, which must be
tested by experiments.
\end{abstract}

\section{Introduction}

While calculations from first principles in QCD are beyond reach,
it is well-known that 
some phenomenological strong interaction models
(such as the MIT bag model) predict metastability and even 
absolute stability of u-d-s quark matter for some ranges of
parameters \cite{Bodmer:1971we,Chin:1979yb,Witten:1984rs,Farhi:1984qu}.
Strange matter metastability could significantly change the properties of
neutron star interiors, and absolute stability (energy per baryon below
that of nuclei) could imply, that 
``neutron stars'' are actually strange stars composed of bulk quark matter
\cite{Witten:1984rs,Haensel:1986qb,Alcock:1986hz,Glendenning:1997wn,Weber:1999qn,Madsen:1998uh,Weber:2004kj}. 
Quark matter in a colour-flavour locked phase and other
colour-superconducting phases, where quarks 
with different colour
and flavour quantum numbers form Cooper pairs with pairing energy
as high as 100~MeV 
is significantly more bound than ordinary 
quark matter, and this increases the likelihood 
that quark matter composed of up, down, and strange quarks rather 
than nuclear matter could be the ground state of bulk hadronic matter
\cite{Alford:1997zt,Rapp:1997zu,Alford:2003kp}.

Finite size effects
(curvature and surface energies) increase the mass per baryon for
small lumps of three flavour quark matter, called strangelets, relative
to bulk quark matter. This and the destabilization in a
high-temperature environment would explain the non-observation of
strangelets in heavy-ion collisions, even if bulk strange quark matter is
stable.

Strangelets may thus exist in at least two possible varieties: 
``Ordinary'' strangelets \cite{Witten:1984rs,Farhi:1984qu}
\ and colour-flavour locked strangelets \cite{Madsen:2001fu}. Both types
consist of almost equal numbers of up, down, and strange quarks so that the
quark charges nearly cancel. The stability of strangelets depends on
the quark masses (especially the heaviest, strange quark) and on the
strong interaction binding (in the MIT bag model characterized by
the bag constant, the one-gluon exchange coupling constant and the
pairing energy). Finite size
effects generally decrease stability at
low baryon numbers relative to bulk strange quark matter, though increased
stability occurs near closed shells. 
``Ordinary'' strangelets are stable for a restricted range of
parameters, and if quark matter is in a colour-flavour locked state, as seems to
be the case at asymptotically high density, the stability is
improved by several tens of MeV per baryon, other parameters fixed.

The characteristic property of strangelets relative to nuclei is a very
small charge for a given baryon number, $A$, because of the near
cancelation of up, down, and strange quark charges in strangelets.
For a typical strange quark mass of 150~MeV, 
``ordinary'' strangelets have charge
$Z\approx0.1A$ for $A\ll700$ and $Z\approx8A^{1/3}$ for $A\gg700$
\cite{Farhi:1984qu,Berger:1986ps,Heiselberg:1993dc}, 
and colour-flavour locked strangelets have
$Z\approx0.3A^{2/3}$ \cite{Madsen:2001fu}.
But strangelet masses can be very large (as high as
the baryon number of a gravitationally unstable strange star,
$A_{\max}\approx2\times10^{57}$), so the quark charge $Z$ can reach values
much higher than those known for nuclei in spite of the low
charge-to-mass ratio.

Atoms can be fully ionized to the nuclear charge, but strangelet
core charges can be high enough that electron-positron
pairs are created in the vacuum, leaving
electrons to screen the central positive charge.
Screened charges of several thousand are easily reached, and there is no
formal maximum charge, though the screened charge increases only slowly with
unscreened charge for high $Z$ \cite{Madsen:2002iw}.

\section{Strangelet detection at the International Space Station}

The Alpha Magnetic Spectrometer AMS-02 on
the International Space Station \cite{AMS:2004}
is an approximately 0.5~m$^2$~sterad acceptance particle physics detector 
with a superconducting magnet providing a 0.86 Tesla dipole field. It is
equipped with a transition radiation detector, silicon trackers, time of
flight detectors, anticoincidence counters, a ring imaging Cherenkov
counter, and an electromagnetic calorimeter, and it
will analyze the flux
of cosmic ray nuclei and particles in unprecedented detail for at least three
years following deployment in 2007 or 2008. Besides all the other
interesting physics to be carried out, it will be sensitive to
strangelets in a wide range of mass and charge, for rigidities in the
GV--TV regime \cite{Sandweiss:2004bu}.
A prototype AMS-01 was successfully flown on the 
Space Shuttle mission STS-91 in 1998 \cite{Aguilar:2002ad}.
Therefore it is of interest to know the 
rigidity spectrum and total flux of cosmic ray
strangelets \cite{Madsen:2004vw}.

If strange matter is stable, neutron stars are likely to be strange
stars, and cosmic ray strangelets will be created when two
strange stars in a binary system collide after inspiral due to loss of
orbital energy in the form of gravitational radiation.
The galactic coalescence rate for neutron (quark) star binaries has
recently been updated from observations of binary pulsars to
\cite{Kalogera:2003tn}
$83.0_{-66.1}^{+209.1}$ Myr$^{-1}$ 
(95\% confidence), thus of order one collision in our Galaxy every
10.000 years.
Binary inspiral lead to tidal disruption of the stars as they
approach each other before the final collision. During this stage small
fractions of the total mass may be released from the binary system in the form
of strange quark matter. 
Simulations of binary neutron star collisions indicate
the release of $10^{-5}-10^{-2}M_{\odot}$ (depending on orbital and
stellar parameters), where $M_{\odot}$ denotes the solar mass.
No high-resolution simulations of collisions involving
two strange stars have been performed, and the physics in black
hole-strange star collisions, where Newtonian and semirelativistic 
results do exist \cite{Lee:2002nk,Kluzniak:2002dm,Prakash:2003em}, 
is too different from the strange star-strange 
star collision to be of real guidance.
Given the stiffness of the equation of
state for strange quark matter, strange star-strange star collisions should
lie in the low end of the mass release range found for
neutron stars, so a conservative estimate of the galactic production
rate of strangelets is
$10^{-10}M_{\odot}\mathrm{yr}^{-1}$.

The quark matter lumps originally released by tidal forces 
are macroscopic in size \cite{Madsen:2001bw}.
When estimated from a balance between quark matter surface tension and tidal
forces, a typical fragment baryon number is
$A\approx 4\times 10^{38}\sigma_{20}a_{30}^3$, where $\sigma_{20}\approx
1$ is the surface tension in units of 20 MeV/fm$^3$ and $a_{30}$ is the
distance between the stars in units of 30 km. 
But subsequent collisions will lead to fragmentation, and
if the collision energy compensates for the
extra surface and curvature energy needed to make smaller strangelets, 
a significant fraction of the mass released from binary strange star
collisions may end up as strangelets with $A\approx
10^{2}-10^{4}$. However, these values are strongly parameter dependent
\cite{Madsen:2001bw}.

Most of the total flux results for cosmic ray strangelets derived in
\cite{Madsen:2004vw}
are such that values for some given $A$ can be used as a lower limit for
the flux for a fixed total strangelet mass injection if strangelets
have a distribution with baryon numbers below $A$.
There is no reliable way of calculating the actual mass spectrum,
so all strangelets released are assumed to have a single baryon number, $A$.

\begin{figure}[h]
\begin{center}
\includegraphics[height=12cm,angle=270]{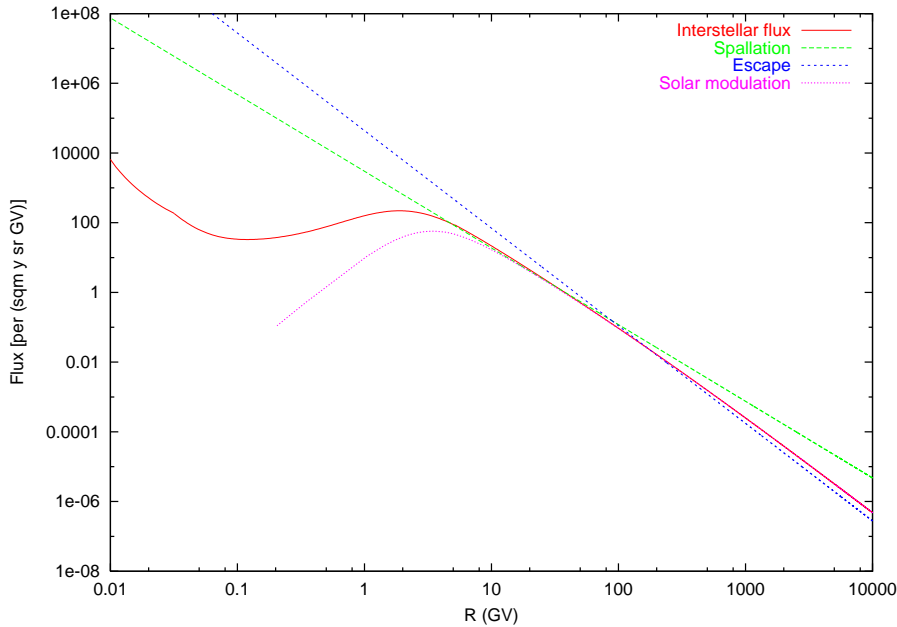}
\caption{The differential flux of strangelets with baryon number 54 and charge 8
(within the range to be probed in an upcoming lunar soil search, as well
as from AMS-02). The interstellar flux curve shows the spectrum before
inclusion of solar modulation, which has been taken into account in the
bottom curve. Curves marked ``Spallation'' and ``Escape'' demonstrate
that certain ranges of rigidity are dominated by strangelet spallation
in the interstellar medium, and escape from the Galaxy, respectively.
Geomagnetic cutoff is not included.}
\label{fig:diffflux}
\end{center}
\end{figure}

Strangelets in many ways behave like cosmic ray nuclei, but with a very 
high $A/Z$-ratio.
Due to the high strangelet rigidity, $R$, at fixed
velocity strangelets are efficiently injected into accelerating shocks,
and most strangelets passed by a
supernova shock will take part in Fermi acceleration. This results
in a source spectrum which is a powerlaw in rigidity.

The timescales for strangelet acceleration, spallation, energy loss, and escape
from the Galaxy are all short compared to the age of the Milky Way Galaxy.
Therefore cosmic ray strangelets can be described
by a steady state distribution, which is a solution to a propagation equation
of the form
\begin{equation}
\frac{dN}{dt}=0
\end{equation}
where $N(E,x,t)dE$ is the number density of strangelets at position $x$ and
time $t$ with energy in the range $[E,E+dE]$. ${dN}/{dt}$
is given by a sum of a source term from supernova acceleration, a
diffusion term, loss terms due to escape from the Galaxy, energy loss,
spallation, and a term to describe reacceleration of strangelets due to
passage of new supernova shock waves.
The various terms and numerical as well as approximate analytical
solutions to the propagation equation are discussed in \cite{Madsen:2004vw}; 
here we
limit ourselves to a few numerical examples shown in the figures, and an
analytic estimate for the total flux.

\begin{figure}[h]
\begin{center}
\includegraphics[height=12cm,angle=270]{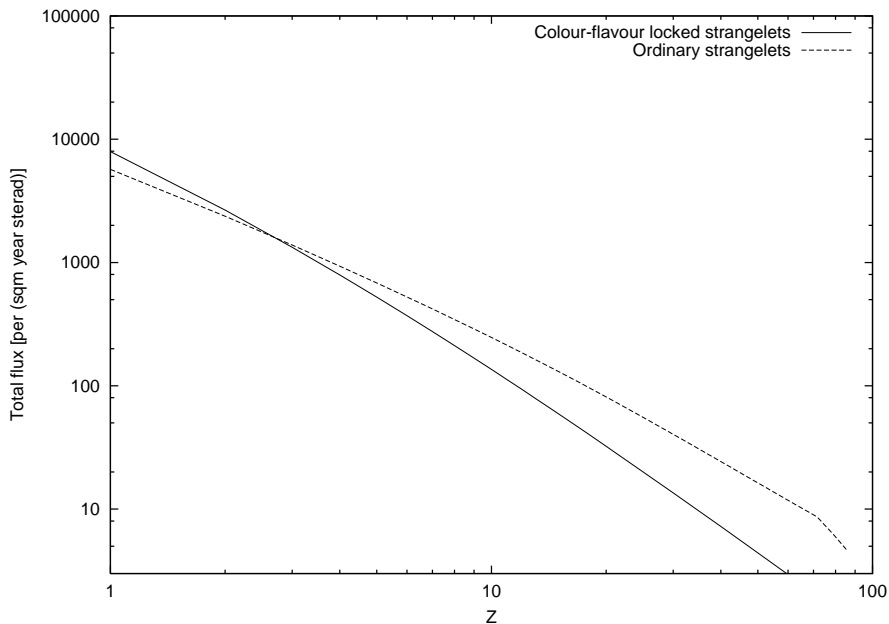}
\caption{The total flux of ordinary and 
colour-flavour locked strangelets as a function
of their total charge for normalization parameter $\Lambda=1$. 
Solar modulation but not geomagnetic cutoff has
been included.}
\label{fig:totflux}
\end{center}
\end{figure}

As discussed in \cite{Madsen:2004vw} solar modulation (the interaction
of cosmic ray strangelets with the outflowing solar wind)
effectively cuts off the strangelet flux at rigidities of order $R_{\mathrm{SM}%
}\approx(A/Z)^{1/2}$GV. It turns out, that 
the strangelet flux is governed by spallation at the cutoff rigidity,
i.e. strangelet destruction in inelastic collisions with interstellar nuclei is
the most important process for galactic strangelets at these rigidities. 
This allows an estimate for the total flux hitting
the Moon or Earth given by
\begin{equation}
F_{\mathrm{total}}\approx2\times10^{5}\mathrm{m}^{-2}\mathrm{yr}^{-1}
\mathrm{sterad}
^{-1}A^{-0.467}Z^{-1.2}\max[R_{\rm{SM}},R_{\rm{GC}}]^{-1.2}\Lambda,
\end{equation}
depending on whether solar modulation or geomagnetic cutoff dominates
(the cutoff rigidity for AMS in the geomagnetic field, $R_{\rm GC}$, is in
the range from 1--15 GV depending primarily on geomagnetic latitude). 
The uncertainties in the normalization of the flux are hidden in the
parameter $\Lambda$, which is proportional to the galactic production
rate of strangelets, inversely proportional to the density of the
interstellar medium and the effective cosmic ray confinement volume in
the Galaxy, etc. The exact definition of $\Lambda$ is given in
\cite{Madsen:2004vw}. $\Lambda$ is of order unity for typical
choices of parameters, including the strangelet production rate of $10^{-10}$
solar masses per year discussed above. 

In the
case of solar modulation domination (always relevant for the Moon, and
relevant for AMS as long as $R_{\rm{SM}}>R_{\rm{GC}}$) one obtains
\begin{equation}
F_{\rm{total}}\approx2\times10^{5}\mathrm{m}^{-2}\mathrm{yr}^{-1}\mathrm{sterad}
^{-1}A^{-1.067}Z^{-0.6}\Lambda,
\end{equation}
which reproduces the numerical results in Figure 2 to 
within 20\% for $Z>10$ and to within
a factor of $2$ even for small $Z$, where the assumptions of nonrelativistic
strangelets and spallation domination are dubious.

Many uncertainties are involved in the calculation of the strangelet flux,
but it is very interesting, that the predictions in Figure 2
are well within reach of an experiment like AMS-02.

\section{Strangelets as ultra-high energy cosmic rays}

The nature of ultra-high
energy cosmic rays at energies above $10^{20}$eV is one of the most
challenging puzzles in modern physics 
(see \cite{Anchordoqui:2002hs,Stanev:2004kk} for recent reviews). 
Protons and nuclei are responsible for most of the cosmic ray flux
at lower energies, but they cannot easily be accelerated to the highest
energies, and even if they
are, their interactions with photons in the cosmic microwave background
radiation are sufficiently energetic to lead to photo-pion and photo-pair
production, as well as (for nuclei) photo-disintegration.
This leads to the so-called GZK-cutoff in flux 
\cite{Greisen:1966jv,Zatsepin:1966jv}
at energies around $10^{19}$eV for protons, and $10^{20}$eV for iron.
The observed flux
shows no clear cut at these energies (though the number of events is small,
and there is inconsistency between different experiments), with observed
cosmic ray energies as high as $3\times10^{20}$eV.
Numerous models have been suggested for the nature of cosmic rays beyond the
GZK-cutoff, a significant fraction of these involving ``new physics'',
but at present no generally accepted explanation exists 
\cite{Anchordoqui:2002hs,Stanev:2004kk}.

Strangelets can be much more massive and have a much higher charge than
nuclei, while still keeping a
low charge-to-mass ratio. These are exactly the properties required to
solve the acceleration
problem \cite{Madsen:2002iw} and move the GZK-cutoff 
to much higher energies \cite{Rybczynski:2001bw,Madsen:2002iw}.

The maximum relativistic energy of a charged cosmic ray particle accelerated in 
an astrophysical ``accelerator'' is given by
$E_{\mathrm{max}}=ZR_{\mathrm{max}}$, where the maximal rigidity
is the value of $R$ where the particle
Larmor radius in the magnetic field (which is proportional to $R$) exceeds the
size of the accelerator. 
Thus, higher charge means higher maximal energy.
For typical cosmic ray nuclei, $Z\leq26$,
and this makes it essentially impossible to reach the highest
cosmic ray energies observed by means of astrophysical source acceleration of
protons or nuclei, given the astrophysical source limitations on
$R_{\mathrm{max}}$. 

But much higher charge
and therefore maximal acceleration energy can be achieved for strangelets
\cite{Madsen:2002iw}.

The average photon energy in the 3K cosmic microwave background radiation is a
mere $E_{\mathrm{3K}}\approx7\times10^{-4}$eV. But an ultra-high energy cosmic
ray proton with a Lorentz-factor in excess of $m_{\pi}/E_{\mathrm{3K}}
\approx10^{11}$ will reach the threshold for the processes $p+\gamma
\rightarrow\pi+\mathrm{nucleon}$, leading to significant energy loss and a
drop in cosmic ray flux at proton energies above $10^{19}$eV.
Photo-disintegration
of nuclei has a threshold of only 10MeV rather than $m_{\pi}$ in the rest
frame of the nucleus, so the cosmic ray flux of nuclei should drop at a
Lorentz factor of order $10^{10}$, corresponding to $E_{\mathrm{GZK}}%
\approx10^{19}A$eV (detailed calculations result in a drop at energies
somewhat lower than found by these simple estimates).

But since the threshold energy is proportional to $A$, high-mass strangelets
are not influenced by these processes at the energies of interest, and
therefore the GZK-cutoff for strangelets moves to energies much larger
than the maximum energy cosmic rays observed to date
\cite{Rybczynski:2001bw,Madsen:2002iw}.

Because of their higher charge, strangelets have larger gyro-radii in
the galactic magnetic field than protons or nuclei. Therefore a
prediction of the strangelet model is, that the distribution of arrival
directions on the sky should become more isotropic for strangelets than
for nuclei \cite{Madsen:2002iw}.

\section{Are stable strangelets ruled out?}

Balberg \cite{Balberg:2004aa} has argued that  
a galactic flux of ultrahigh energy strangelets is basically ruled out
because they would trigger 
transformation of all neutron stars into strange quark matter stars. 
He argued, that all neutron stars can not be strange stars, 
and therefore finds the scenario in \cite{Madsen:2002iw} unlikely.

As I shall briefly discuss below, neither of these arguments are
necessarily
correct, but they revive an old discussion about the (im)possibility of
stable strange quark matter which it seems worthwhile to address.

The first assumption in \cite{Balberg:2004aa} 
neglects the fact that strangelets
at the relevant energies will be destroyed in collisions with the stars
they are supposed to transform. 
Strangelet fragmentation will occur if
the total energy added in inelastic collisions with
nuclei exceeds the strangelet binding energy, which can be
some tens of MeV per baryon. While the details of
strangelet fragmentation remains to be studied (much of the necessary
input physics is poorly known) order-of-magnitude estimates
demonstrate that ultrahigh energy strangelets
will be destroyed in collisions rather than serve as seeds to
transform neutron stars into strange stars \cite{Madsen:2004an}.

In contrast to the extremely high energy cosmic ray strangelets
treated in \cite{Madsen:2002iw},
even a small flux of strangelets at low energies 
would be able to convert all neutron stars
into strange stars. This was discussed in detail in 
\cite{Madsen:1989pg,Friedman:1990qz}. 
At that time it was argued that this ruled out
the hypothesis of stable strange quark matter and strangelets, because
some properties of pulsars (especially the sudden spin-up events called
glitches) seemed
inconsistent with strange star properties. Balberg \cite{Balberg:2004aa}
revives these arguments and lists a set of such
properties including glitches, r-mode instabilities and cooling.
However at the current level of understanding one should not rule
out strange stars on these grounds. The strange star glitch problem
\cite{Alpar:1987vk} has been shown to be marginally consistent with ordinary
strange stars with nuclear crusts \cite{webglen}, and may also find an
explanation in the crystalline phases discovered in 
colour-superconducting
quark matter \cite{Alford:2000ze}. The r-mode instabilities that rule
out the simplest colour-flavour locked strange stars 
\cite{Madsen:1999ci,Madsen:1998qb} could
also be consistent in models with crystalline phases,
and similar counterexamples exist for the other effects mentioned in
\cite{Balberg:2004aa}. It is not known at present 
whether strange stars exist at all, but
it is not ruled out either, that all ``neutron'' stars could be strange
stars.

\section{Conclusion}

A strangelet scenario for ultrahigh energy cosmic rays
provides simple explanations for the acceleration problem and
the GZK-cutoff problem. It is however only one among many models
in the literature. 

There is no convincing evidence either in favour or
against stable strangelets in ordinary cosmic rays. The
flux at GV--TV rigidities could well be in a range that 
can be probed in the upcoming AMS-02
experiment on the International Space Station. There is also hope for
testing the stable strangelet hypothesis in a more restricted mass and
charge range in a coming search in lunar soil, where strangelets
could have been deposited over timespans of hundreds of millions of
years.

Strange quark matter stability is an exciting possibility. The validity
of the idea can not be decided on the basis of theory alone.
Fortunately, experimental tests involving cosmic rays may settle the issue
within a few years.

\ack
This work was supported by The Danish Natural Science Research Council.

\end{document}